\documentstyle[12pt]{article}

\begin{document}

\title{\hfill  UGI-96-03\\[1cm]
Dilepton anisotropy from $p + Be$ and $Ca + Ca$
collisions at BEVALAC energies
\footnote{Supported by BMFT and GSI Darmstadt} \vspace*{5mm} \\}
\author{E.L. Bratkovskaya\thanks{
Permanent address: Bogoliubov Laboratory of Theoretical Physics,
Joint Institute for Nuclear Research, 141980 Dubna, Moscow Region, Russia},
W. Cassing, and U. Mosel\vspace*{2mm} \\
\small \em Institut f\"ur Theoretische Physik, Universit\"at Giessen,
D-35392 Giessen, Germany}
\date{}
\maketitle

\begin{abstract}
A full calculation of lepton-pair angular characteristics is carried
out for $e^+e^-$ pairs created in $p + Be$ and $Ca + Ca$ collisions
from 1.0 to 2.1 GeV/A. It is demonstrated that the dilepton decay
anisotropy depends sensitively on the different sources and may
be used for their disentangling. Due to the dominance of the
$\eta$-and $\Delta$-Dalitz decays and only a small anisotropy
coefficient for $\pi^+\pi^-$ annihilation, the expected anisotropy
coefficients show a decrease with invariant mass of the dilepton pair
and change only moderately when comparing $p + Be$ and $Ca +
Ca$ reactions at the same bombarding energy.
\end{abstract}

\newpage
Dileptons are quite attractive electromagnetic signals since they
provide almost direct information on the hot and dense nuclear phase in
heavy-ion collisions at BEVALAC/SIS and SPS energies
\cite{ro88,na89,ro89,CERES}.  The information carried out by leptons
may tell us not only about the interaction dynamics of colliding
nuclei, but also on properties of hadrons in the nuclear environment or
on a possible phase transition of hadrons into a quark-gluon plasma
(cf. \cite{Mosel91}).  However, there are a lot of hadronic sources for
dileptons because the electromagnetic field couples to all charges and
magnetic moments. In particular, in hadron-hadron collisions, the
$e^+e^-$ pairs are created due to the electromagnetic decay of
time-like virtual photons which can result from the bremsstrahlung
process or from the decay of baryonic and mesonic resonances including
the direct conversion of vector mesons into virtual photons in
accordance with the vector dominance hypothesis.  In the nuclear
medium, the properties of these sources may be modified and it is thus
very desirable to have experimental observables which allow to
disentangle the various channels of dilepton production.

Recently, we have proposed to use lepton pair angular distributions for
a distinction between different  sources \cite{BTT94,BSCMTT94}.
Indeed, it could be shown that due to the spin alignment of the virtual
photon and the spins of the colliding or decaying hadrons, this {\em
lepton decay anisotropy} turns out to be quite sensitive to the
specific production channel (cf. also
\cite{Hag95,GulTitov95}).  Whereas in our previous work
\cite{BTT94,BSCMTT94,BCMTTT95} we have considered the dilepton
anisotropy for elementary nucleon-nucleon or $pd$ collisions, we now
evaluate this new observable for the first time for proton-nucleus and
nucleus-nucleus collisions from 1 to 2~GeV/A bombarding energy, where
the calculated inclusive dilepton spectra can be controlled in
comparison to the DLS data~\cite{ro88,na89,ro89}.

The dynamical evolution of the proton-nucleus or nucleus-nucleus
collision is described by a transport equation of the
Boltzmann-Uehling-Uhlenbeck type evolving phase-space distribution
functions for nucleons, $\Delta$'s, $N^*(1440)$'s, $N^*(1535)$'s, pions
and $\eta$'s with their isospin degrees of freedom.  The details of the
model are discussed in Refs.~\cite{Wolf90,Wolf_eta92}; here we use the
same prescriptions and include additionally the production channels
from the direct decay of the vector mesons $\rho, \omega,$ and $\Phi$
as well as the Dalitz-decay of the $\omega$-meson \cite{Cass95}
employing the formfactors from Landsberg \cite{Landsberg}.

In  Fig.~\ref{fig1} we present the calculated dilepton invariant mass
spectra $d\sigma/dM$ for $p+{}^9Be$ and ${}^{40}Ca+{}^{40}Ca$
collisions at bombarding energies from 1 to 2.1~GeV/A and compare them
with the experimental data of the DLS
collaboration~\cite{ro88,na89,ro89} including the DLS acceptance filter
as well a mass resolution of 50 MeV.  In the 'cocktail' plot the
$\Delta$ Dalitz decay (labeled by '$\Delta$'), $N^*$ Dalitz decay
('$N^*$'), proton--neutron bremsstrahlung ('$pn$'), $\pi N$
bremsstrahlung ('$\pi N$') (all without VDM formfactor), $\omega\to\pi
e^+e^-$ Dalitz decay, $\eta$ Dalitz decay ('$\eta$'), the pion
 annihilation channel ('$\pi\pi$') as well as the direct decay of the
vector mesons ('$\rho,\omega,\Phi$') are separated explicitely while
the solid curves (denoted by '$all$') represent the sum of all sources.
As seen from  Fig.~\ref{fig1} the dilepton cross sections  agree
reasonably well with the experimental data as in \cite{Wolf_eta92}.  We
will use the same cross sections, however, without experimental DLS
filter for the calculation of the anisotropy coefficients.

In order to characterize the dilepton decay anisotropy we have introduced in
\cite{BTT94,BSCMTT94} the anisotropy coefficient $B$ which
allows to characterize  the angular distribution of dileptons
created in a hadron-hadron ($h+h$) reaction via
\begin{eqnarray}
S_i(M,\theta ) \equiv {d\sigma_i^{hh}\over dM d\cos\theta_{hh}}
= A_i^{hh} (1 + B_i^{hh} \cos^2\theta_{hh}).
\label{in1}\end{eqnarray}
In Eq. (1) $\theta_{hh}$ denotes the angle of the electron
momentum $\vec l_-^*$ -- measured
in the dilepton center-of-mass system ($\vec q^* \equiv \vec l_-^* + \vec
l_+^* = 0$) -- with respect to the velocity of the $h + h$ c.m.s. (denoted
by $\vec v^{hh}$) relative to the dilepton c.m.s., i.e.
$\vec v^{hh}= - {\vec q}^{hh}/q_0^{hh} = - \vec v_q^{hh} $:
$\cos \theta_{hh}=(\widehat{\vec l_-^*, \vec v_q^{hh}})$.
Here, the momentum and energy of virtual photon ($q_0^{hh}, \ \vec q^{hh}$)
are defined in the c.m.s. of the $h+h$ system while $M$ is the
invariant mass of a lepton pair, $(M^2 = q_0^2 -\vec q^2) $.
$B_i^{hh}$ describes the anisotropy while $A_i^{hh}$ determins magnitude
of the cross section.

The total differential cross section for $h+h$ collisions now
can be represented as a sum of the differential cross sections for
all channels,
\begin{eqnarray}
{d\sigma^{hh} \over dM d \cos\theta_{hh}} = \sum\limits_{i=channel}
{d\sigma_i^{hh}\over dM d\cos\theta_{hh}} =
A^{hh}(M) (1+B^{hh}(M)\cos^2\theta_{hh}),
\label{Ssum}\end{eqnarray}
which leads to the total anisotropy coefficient:
\begin{eqnarray}
 B^{hh}(M) = \sum\limits_{i=channel} <B_i^{hh}(M)>,  \hspace*{1cm}
 <B_i^{hh}(M)> = {\displaystyle  {\displaystyle d\sigma_i^{hh}\over dM} \cdot
{\displaystyle B_i^{hh}\over 1+{\displaystyle 1\over 3} B_i^{hh}} \over
\displaystyle \sum\limits_i {\displaystyle d\sigma_i^{hh}\over dM} \cdot
{\displaystyle 1\over 1+{\displaystyle 1\over 3} B_i^{hh}}},
\label{B_isum}\end{eqnarray}
where the special weighting factors originate from the necessary
angle-integrations.
Thus, the anisotropy coefficient $B^{AB}$ for $A+B$ reactions is the
sum of the ``weighted'' anisotropy coefficients ($<B_i^{AB}>$) for each
channel $i$ obtained by means of the convolution of $B_i^{AB}$ with the
corresponding invariant mass distribution (cf. \cite{BCMTTT95}).

For heavy-ion reactions the situation becomes more complicated
due to the nuclear dynamics and the explicit time evolution of the
interacting system.  Here we start from the point that the form of the
angular distribution for all 'elementary' interactions $a+b$, that
occur in the nucleus-nucleus reaction $A+B$, are known.  For this aim we
employ the results of our previous works \cite{BTT94,BSCMTT94,BCMTTT95}
where the anisotropy coefficients for the $pp$, and $pn$
bremsstrahlung, $NN\to \Delta \to NN e^+e^-$ Dalitz decay, $\eta$
Dalitz decay and $\pi\pi$ annihilation channels were calculated
explicitly in the hadron-hadron center-of-mass system $a+b$.

The differential angular distribution in elementary $a+b$ collisions
-- before averaging over the momentum $\vec q^{ab}$ -- can be represented as:
\begin{eqnarray}
{d \sigma^{ab}_i\over dM \ d\vec q^{ab} \ d \cos\theta_{ab}} = A_i^{ab}
\left( 1 + B_i^{ab} \cos^2\theta_{ab} \right),
\label{dsdmq}\end{eqnarray}
where the coefficient $B_i^{ab}$ for the elementary process $a+b$
is a function of the dilepton mass $M$, the masses $m_a, m_b$  and the 
initial invariant energy $\sqrt{s}$ of the hadrons involved in the reaction:
$B_i^{ab}=B_i^{ab}(M,\sqrt{s};m_a,m_b)$.
It is important to note that we know the 'elementary' coefficient only
in the $a+b$ system.  In nucleus-nucleus collisions, however, the
direction ${\vec v}_q^{ab}={\vec q}^{ab}/q_0^{ab}$ is changed in each
elementary $a+b$ collision.  Thus in order to define an anisotropy
coefficient in the latter situation we have to perform an angular
transformation from the elementary c.m.s.  ($\theta_{ab}$) to the
c.m.s. of the colliding nuclei ($\theta_{AB}$):
\begin{eqnarray}
\cos\theta_{ab}=\cos\theta_{AB} \cos\theta_q + \sin\theta_{AB}
\sin\theta_q \cos(\varphi_{AB} -\varphi_q),
\label{cos}\end{eqnarray}
where $\theta_q$ is the angle between the dilepton c.m.s. velocity
${\vec v}_q^{ab}={\vec q}^{ab}/q_0^{ab}$ in the c.m.s. of $a+b$
and the dilepton c.m.s. velocity ${\vec v}_q^{AB}={\vec q}^{AB}/q_0^{AB}$
in the c.m.s. of the colliding nuclei $A+B$:
$\cos\theta_q = \vec v_q^{ab}\cdot \vec v_q^{AB} / (|\vec v_q^{ab}| \
|\vec v_q^{AB}|)$. Here, the vector $\vec q^{ab}$ is obtained by a
Lorentz transformation of  $\vec q^{AB}$:  $\ \vec q^{ab} = L(\vec V^{ab})
\vec q^{AB}$, where $\vec V^{ab}=(\vec p_a+\vec p_b)/(E_a+E_b)$
is the velocity of the $a+b$ system relative to the $A+B$ system.

Substituting (\ref{cos}) into Eq.(\ref{dsdmq}) and using
$d\Omega_{AB}=d\Omega_{ab}$ we get the respective
distribution in the observable angle $\theta_{AB}$ by integrating over
the azimuthal angle $\varphi_{AB}$:
\begin{eqnarray}
&& {d \sigma^{ab}_i\over dM d \cos\theta_{AB} } \sim \int d\vec q^{AB}  \
\tilde A_i(\theta_q) \ \left(1 + \tilde B_i(\theta_q) \cos^2\theta_{AB}
\right),	\label{dsdmqtrans} \\
&& \tilde A(\theta_q) = 1 + {B_i^{ab}\over 2} \sin^2\theta_q,
\hspace*{5mm}  \tilde B(\theta_q) = {B_i^{ab}\over 2 \tilde A(\theta_q)}
(3 \cos^2\theta_q - 1).\label{tildeAB}
\end{eqnarray}
Obviously, the transformation (\ref{cos}) does not change the quadratic
form in $\cos\theta$ of the angular distribution.
From Eqs.(\ref{in1}) and (\ref{dsdmqtrans}) and the normalization condition
we finally get
\begin{eqnarray}
B_i^{AB}={ \int d\vec q^{AB} \ \tilde A(\theta_q) \ \tilde B(\theta_q) \
\displaystyle {d\sigma_i^{ab}\over dM d\vec q^{AB}}  \over
\int d\vec q^{AB} \ \tilde A(\theta_q) \ \displaystyle {
d\sigma_i^{ab}\over dM d\vec q^{AB}}},
\label{BAB}\end{eqnarray}
where the differential cross sections $d\sigma_i^{ab} / (dM d\vec
q^{AB})$ for the different channels i are taken from the BUU
calculations as described in \cite{Wolf90,Wolf_eta92}. The respective
'reduced' differential cross sections $d\sigma_i^{AB}/dM$ are shown in
Fig.~\ref{fig1} for the systems to be studied below.

	The above definition of the anisotropy coefficient in
nucleus-nucleus collisions is valid for all channels except for
$\pi^+\pi^-$--annihilation because there are only two particles in the
initial and final states.
For this particular reaction we have to use the
angle of the pion momentum with respect to the lepton momentum in the
c.m.s. of the leptons (or pions, which is the same for this channel),
i.e. ${\vec q}^{ab}=\vec p_a+\vec p_b = \vec l_+ +\vec l_- = 0$.
As was shown in Ref.~\cite{BTT94}, the
elementary anisotropy coefficient in the $\pi \pi$ c.m.s. then is
$B_{\pi^+\pi^-} = -1$.
In heavy-ion collisions we can use the same definition for the polar angle
as for the other channels because we can reconstruct the direction of
$\vec q^{AB}$ in the c.m.s. of the colliding nuclei $A+B$.  Furthermore, we
performed the same angular transformation from the c.m.s. of the pions
$a+b$ to the c.m.s. of the nuclei $A+B$ as in Eq.(\ref{cos}) with the
replacement $\theta_q \to \theta_\pi$.  Here, $\theta_\pi$ is the angle
between the pion momentum $\vec p_a$ in the c.m.s. of $a+b$ and the vector
$\vec v_q^{AB}={\vec q}^{AB}/q_0^{AB}$.  Following the same angular
integration as for the previous cases we end up with the expression for
the anisotropy coefficient for $\pi^+ \pi^-$ annihilation:
\begin{eqnarray}
B_{\pi\pi}^{AB}={ \int d\cos\theta_\pi \ \tilde A(\theta_\pi) \
\tilde B(\theta_\pi) \ W(\cos\theta_\pi)  \over
\int d\cos\theta_\pi \ \tilde A(\theta_\pi) \ W(\cos\theta_\pi)}.
\label{Bpi}\end{eqnarray}
The quantities $\tilde A(\theta_\pi), \tilde B(\theta_\pi)$ are defined by
Eq.(\ref{tildeAB}) using $B_{\pi\pi}^{ab}=-1$ while
$\cos\theta_\pi$ can be computed by the scalar products of the
four momenta of pions $p_a, p_b$ and colliding nuclei $P_A, P_B$,
\begin{eqnarray}
\cos\theta_\pi= {(p_a -p_b).(P_A+P_B) \ M \over \sqrt{(M^2-4m_\pi^2) \
\left( \ \left[(p_a+p_b).(P_A+P_B)\right]^2 - (P_A+P_B)^2 \ M^2 \right)}},
\label{cosinv}\end{eqnarray}
with $M^2=s_{ab}=(p_a+p_b)^2$ following O.V.~Teryaev~\cite{Oleg}.

A closer look at Eq.(\ref{Bpi}) shows that $B_{\pi\pi}^{AB}$ is a
constant; its absolute value follows from the pion angular distribution
$W(\cos\theta_\pi)$.  In case of an isotropic angular distribution,
$W(\cos\theta_\pi)=const$, the anisotropy coefficient
$B_{\pi\pi}^{AB}=0$ according to Eq.(\ref{Bpi}).  Moreover, in the BUU
calculation the angular distribution $W(\cos\theta_\pi)$ is also a
function of time $t$ due to the dynamical evolution of the system.
Technically we define the distribution $W(t)$ at time $t$ (i.e. in the
time interval $[t- \Delta t/2:  t+ \Delta t/2]$)  as the ratio of the
number of pions with fixed $\cos\theta_\pi$ produced in the above time
interval to the total number of
pions produced in the $A + B$ reaction, i.e.  $W(t,\cos\theta_\pi) =
\dot{N}(t,\cos\theta_\pi)/N_{tot}$.

In order to demonstrate the pion angular anisotropy we show in
Fig.~\ref{fig2} (l.h.s.) the pion angular distribution
$W(t,\cos\theta_\pi)$ for $p+{}^9Be$ at 2.1~GeV and
${}^{40}Ca+{}^{40}Ca$ at 1.0 and 2.0~GeV/A at those times $t$ when the
maximal number of pions $N(t)$ was produced during the interval $\Delta t$;
the respective $\pi$-production rate $\dot{N}(t)/N_{tot}$ is
illustrated in the r.h.s. of
Fig.~\ref{fig2} for the same reactions. Due to the low number of pions
produced for $p+{}^9Be$ the angular distribution suffers from low
statistics and the error bars -- which result from different runs with
8000 testparticles per nucleon -- are large. For the further
analysis we have used the dotted line in Fig.~\ref{fig2} which
represents a fit to the angular distribution. On the other hand, for
${}^{40}Ca+{}^{40}Ca$ collisions the statistics is good enough and we
show only fits to the 'numerical' data in terms of the solid and
dash-dotted lines, respectively.  The anisotropy of $W(\cos\theta_\pi)$
is most pronounced for $Ca + Ca$ at 1 GeV/A, because the probability
for $\pi$'s with transverse momentum is larger then with longitudinal
momentum due to a stronger pion absorption in beam (longitudinal)
direction. At 2 GeV/A, the pion distribution becomes more isotropic in
line with a 'pionic fireball' scenario. Since pion absorption effects
are only small for $p + Be$, the resulting angular distribution shows
only a very modest anisotropy.

Before going over to the calculation of the anisotropy coefficient for
nucleus-nucleus collisions one has to take into account that resonances
($\Delta, N^*$) can be created in quite different elementary channels
than in $pN$ or $pd$ reactions.  For example, $\Delta$ production via the
$\pi N \to \Delta$ channel becomes quite important; an elementary
channel for which the dilepton anisotropy has not been computed so far.
We thus have calculated the $e^+e^-$ anisotropy using the same
vertices, delta-propagator, coupling constants and formfactors as in
Refs.~\cite{BSCMTT94,Schafer94}. Since this evaluation is straight
forward, we do not present the details here.

The results of our microscopic calculation for the anisotropy
coefficient for the $\pi + N \to \Delta \to e^+e^- + X$ channel are
displayed in Fig.~\ref{fig3}.  The anisotropy $B_{\pi N \to \Delta}$
is a function of the dilepton invariant mass $M$ and the
invariant energy of the interacting particles $s=(p_\pi + p_N)^2 \equiv
M_\Delta^2$. For small $\sqrt{s}$ only deltas with $M_\Delta \approx
M_{\Delta 0} =1.232$~GeV appear and the coefficient $B_{\pi N \to
\Delta} \to 1$ as expected for the Dalitz decay of a free delta
\cite{BTT94}. With increasing $\sqrt{s}$ more energetic deltas can be
created and (for fixed $M$) the phase-space for the final nucleon and
virtual photon increases leading to a decrease of $B_{\pi N \to \Delta}$.
We note that we do not take into account the Dalitz decay of
the higher nucleon resonances in the further calculations because their
statistical weight is too low.

Fig.~\ref{fig4} finally shows the computed weighted anisotropy coefficients
$<B_i(M)>$ for $p+{}^9Be$ and ${}^{40}Ca+{}^{40}Ca$ collisions at the
bombarding energies from 1 to 2.1~GeV/A. The main contributions arise
from the $\eta$ and $\Delta$ Dalitz decays due to their large
'elementary' anisotropy coefficients and cross sections (cf.
Fig.~\ref{fig1}), respectively. The contribution from $pn$
bremsstrahlung is practically zero at all energies due to a smaller
'elementary' anisotropy coefficient and due to a lower cross section as
well.  The weighted coefficient from $\pi^+\pi^-$ annihilation is
rather small ($\approx$ 0.1) even for the $Ca  + Ca$ reactions and
decreases for $M \geq m_\rho$ due to the threshold behaviour of the
cross section. However, compared to $<B_{\pi^+\pi^-}(M)> $ for
$p+{}^9Be$, where pion annihilation is very low, a clear (but moderate)
enhancement can be extracted.

	The contributions of the further channels ($N^*$, $\pi N$
bremsstrahlung, $\omega\to \pi^0 e^+e^-$) are also negligible due to
their small cross sections (cf. Fig.~\ref{fig1}). The cross section
from direct decays of the vector mesons becomes compatible with pion
annihilation for $M \approx m_\rho$ for $p+{}^9Be$ at 2.1 GeV, but the
anisotropy coefficient for the 'free' vector meson in the vacuum is
zero \cite{GaleK}. We do not discuss here a possible modification of
the $\rho$-meson properties in the medium that might lead to
non-isotropic angular distributions of dileptons because for $p+{}^9Be$
at 2.1~GeV one cannot reach sufficiently high baryon or pion densities.

On the other hand, for ${}^{40}Ca+{}^{40}Ca$ collisions  at 2.0~GeV and
for $p+{}^9Be$ at 2.1~GeV we observe a clear anisotropy from the $\eta$
Dalitz decay, while for $p+{}^9Be$ at 1~GeV the $\Delta$ Dalitz decay
gives the main contribution at small invariant mass due to a dominant
$\Delta$ cross section; the $\eta$ coefficient increases at $M$ from
0.4 to 0.5~GeV for the same reason.

For clarity we briefly discuss  the steps to
'extract' the anisotropy coefficient from experimental data.
In dilepton experiments the four-momenta of leptons
in the c.m.s. of the nuclei $A+B$ (or in the laboratory frame, which
is connected with the c.m.s. by a simple Lorentz transformation)
are measured --
$(\varepsilon_-^{AB},\vec l_-^{AB}), \ (\varepsilon_+^{AB},\vec l_+^{AB})$.
The four momentum $q=(q_0^{AB}, \vec q^{AB})$ of the virtual photon
in the $A+B$ c.m.s. then is given by:
$q_0^{AB}= \varepsilon_-^{AB} + \varepsilon_+^{AB}, \
\vec q^{AB} =\vec l_-^{AB} + \vec l_+^{AB}$.
The angle $\theta_{AB}$ has to be computed via:
\begin{eqnarray}
\cos\theta_{AB}={|\vec q^{AB}|-q_0^{AB} \cos\theta_{ql}\over
|\vec q^{AB}|\cos\theta_{ql}-q_0^{AB}}, \label{cosdef}
\end{eqnarray}
where $\theta_{ql}$ is the angle between the $\vec q^{AB}$ and
$\vec l_-^{AB}$, i.e. $\cos\theta_{ql} = \vec l_-^{AB} \cdot \vec q^{AB} /
(|\vec l_-^{AB}|\ |\vec q^{AB}|)$.

In order to extract the anisotropy coefficient one has to count
the number of dilepton events $N(M, \cos\theta_{AB})$
with fixed $\cos\theta_{AB}$ and invariant mass $M$,
The anisotropy coefficient for the invariant mass $M$ then is
simply given by:
\begin{eqnarray}
B^{AB}(M) = {N (M, \cos\theta_{AB}=1)\over N (M, \cos\theta_{AB}=0)} -1.
\label{Bexp}\end{eqnarray}

Thus, summarizing, the calculated anisotropy coefficients for $p + Be$
and $Ca + Ca$ collisions support our suggestion in
Refs.~\cite{BTT94,BSCMTT94,BCMTTT95} that the dilepton decay anisotropy
may serve as an additional observable to decompose the dilepton spectra
into the various sources. Since the anisotropy  vanishes
in a hadronic fireball scenario our present results provide valuable
information about the nonequilibrium stage of the reactions.

\vspace*{5mm}
We gratefully acknowledge many helpful discussions with Gy. Wolf,
S.S. Shimanskij, S. Teis, O.V. Teryaev, A.I. Titov, and D.V. Toneev.

\newpage
\section*{Figure captions}

\begin{figure}[h]
\caption{
The dilepton invariant mass spectra $d\sigma/dM$ from $p+{}^9Be$
and ${}^{40}Ca+{}^{40}Ca$ collisions at bombarding energies from 1 to
2.1~GeV/A in comparison to the experimental data~\protect\cite{ro88,na89,ro89}.
The ``$\eta$'' denotes the contribution of the $\eta$--channel, the
``$\Delta$'' labels the contribution of the $\Delta$ Dalitz decay,
``$N^*$'' the $N^*$ Dalitz decay, ``$\omega\to\pi e^+e^-$" is the
$\omega\to\pi e^+e^-$ Dalitz decay, ``$pn$''  the proton--neutron
bremsstrahlung, ``$\pi N$'' the $\pi N$ bremsstrahlung, ``$\pi\pi$''
the pion annihilation channel, while ``$\rho,\omega,\Phi$'' is the direct
decay of vector mesons .  The solid curves (denoted by ``$all$'') show
the sum of all sources.}
\label{fig1}

\caption{
The pion angular distribution $W(t,\cos\theta_\pi)$
for $p+{}^9Be$ at 2.1~GeV and ${}^{40}Ca+{}^{40}Ca$ at 1.0, 2.0~GeV and
time $t$ (l.h.s.) where the maximal number of pions $N(t)$ was produced.
The pion production rate ${\dot N}(t)/N_{tot}$ is shown
in the r.h.s. of the
figure. }
\label{fig2}

\caption{
The results of our calculation for the anisotropy coefficient of the
$\pi + N \to \Delta \to e^+e^- + X$ channel at
$M_\Delta=\protect\sqrt{s}$ from 1.232 to 1.8~GeV. }
\label{fig3}

\caption{
The weighted anisotropy coefficients $<B_i(M)>$ for $p+{}^9Be$ and
${}^{40}Ca+{}^{40}Ca$ collisions at bombarding energies from 1 to
2.1~GeV/A.  The notation is the same as in Fig.~\protect\ref{fig1}. }
\label{fig4}

\end{figure}

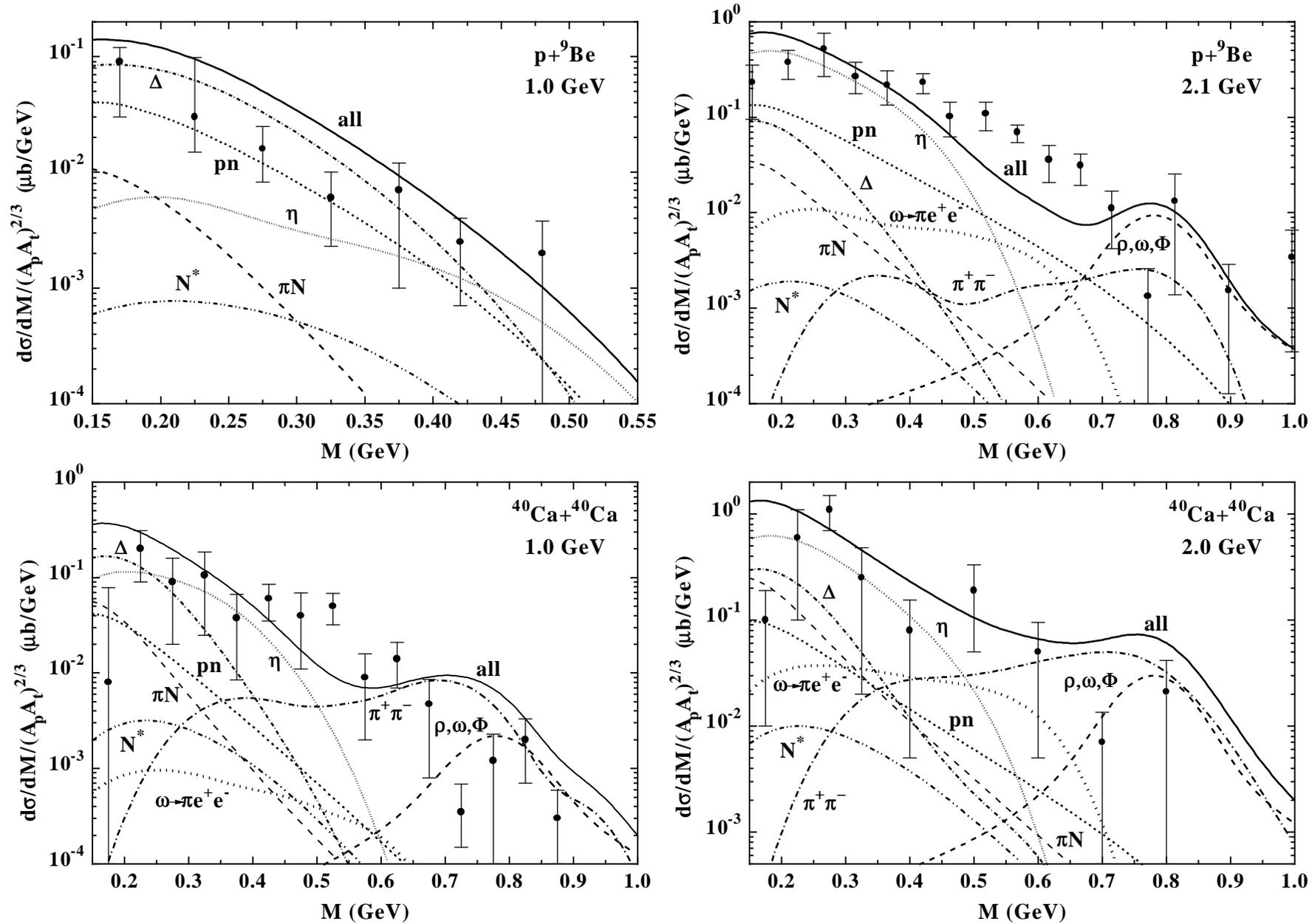
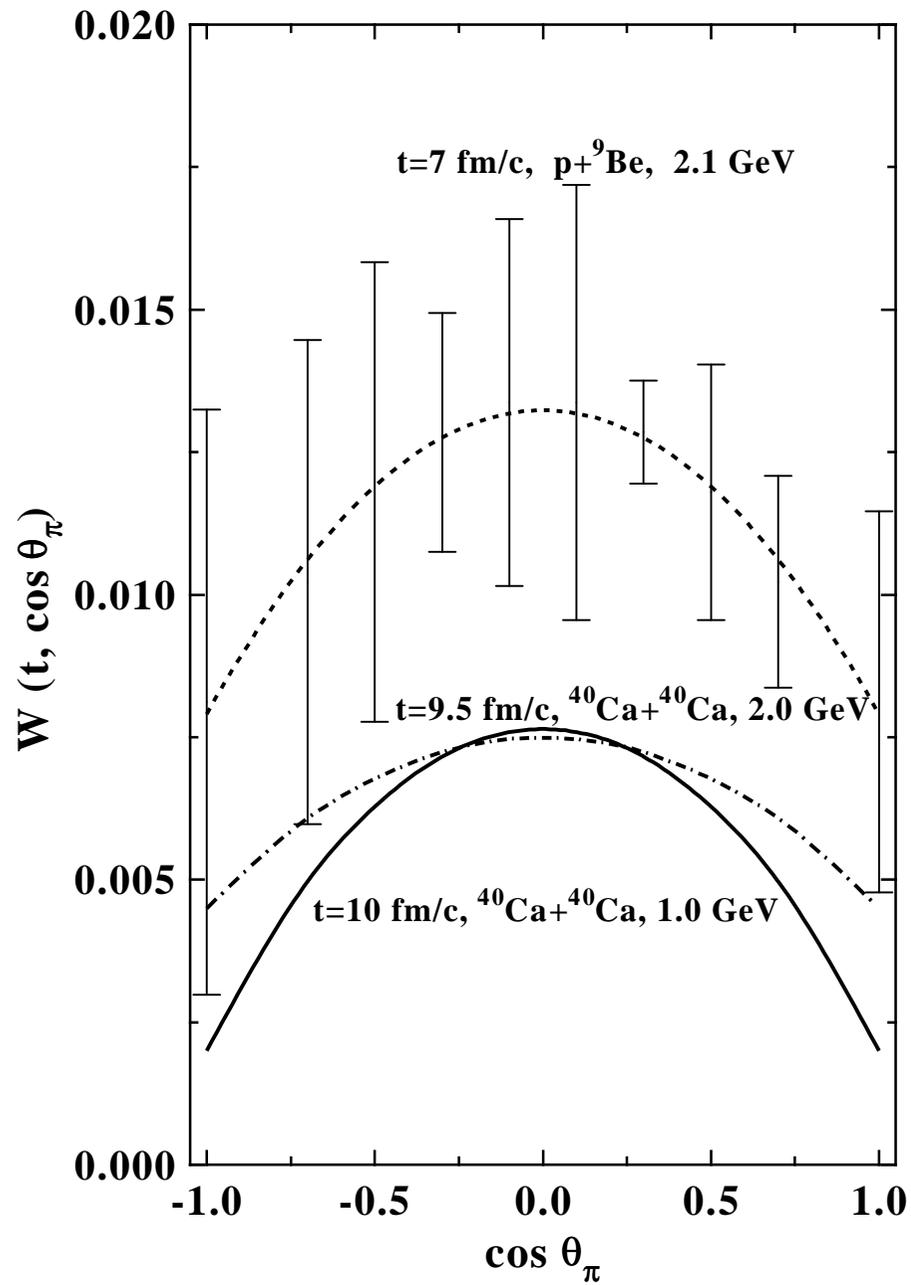
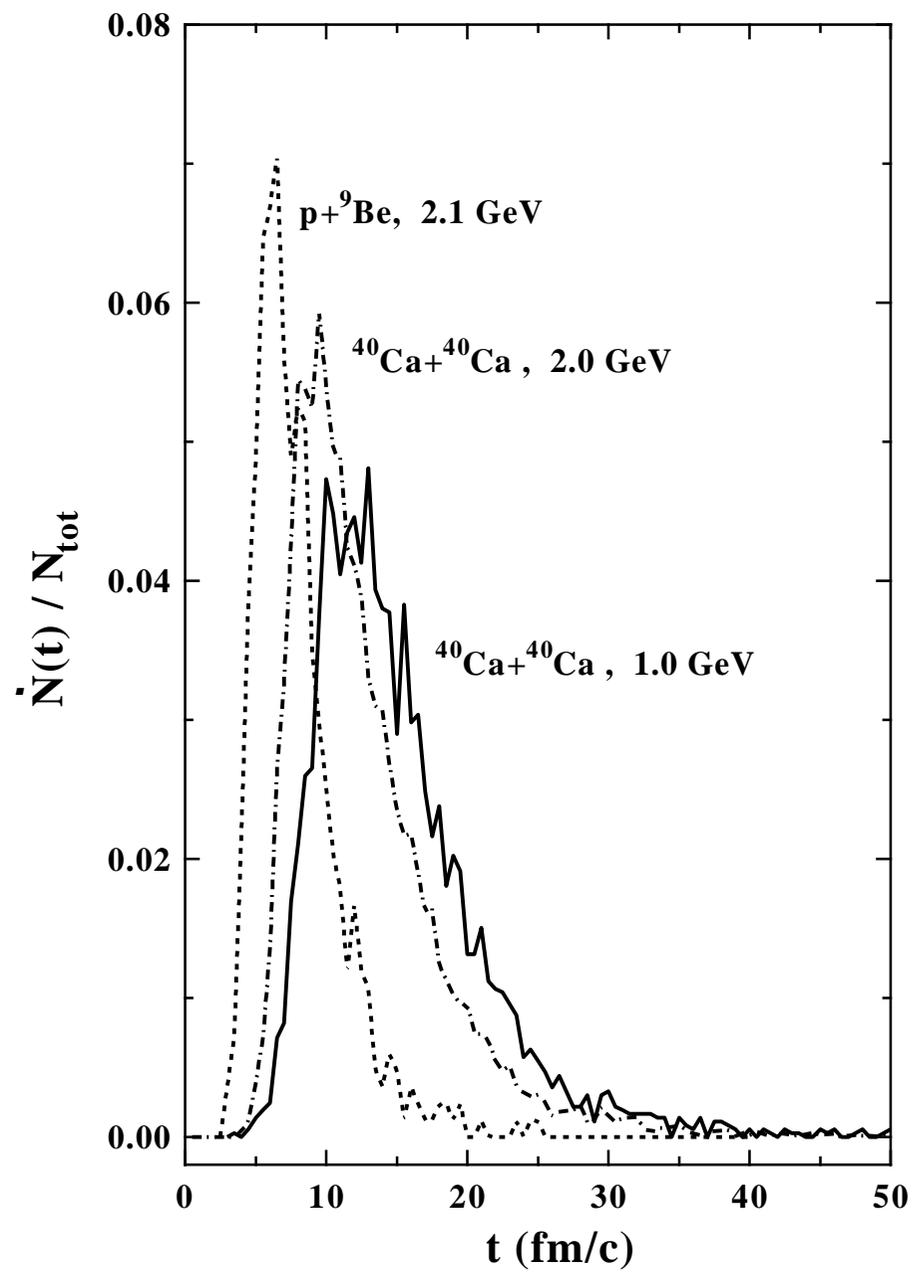
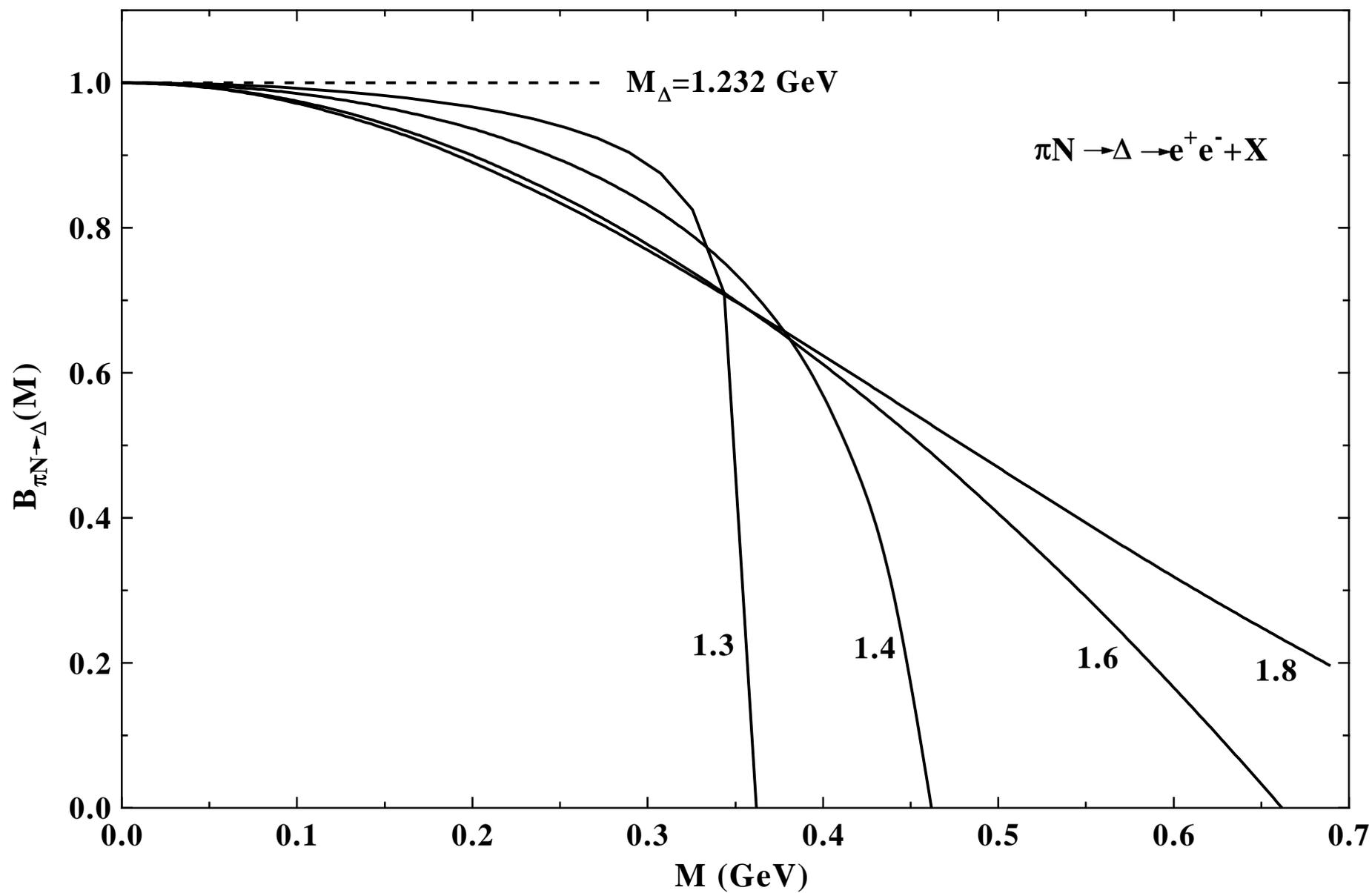
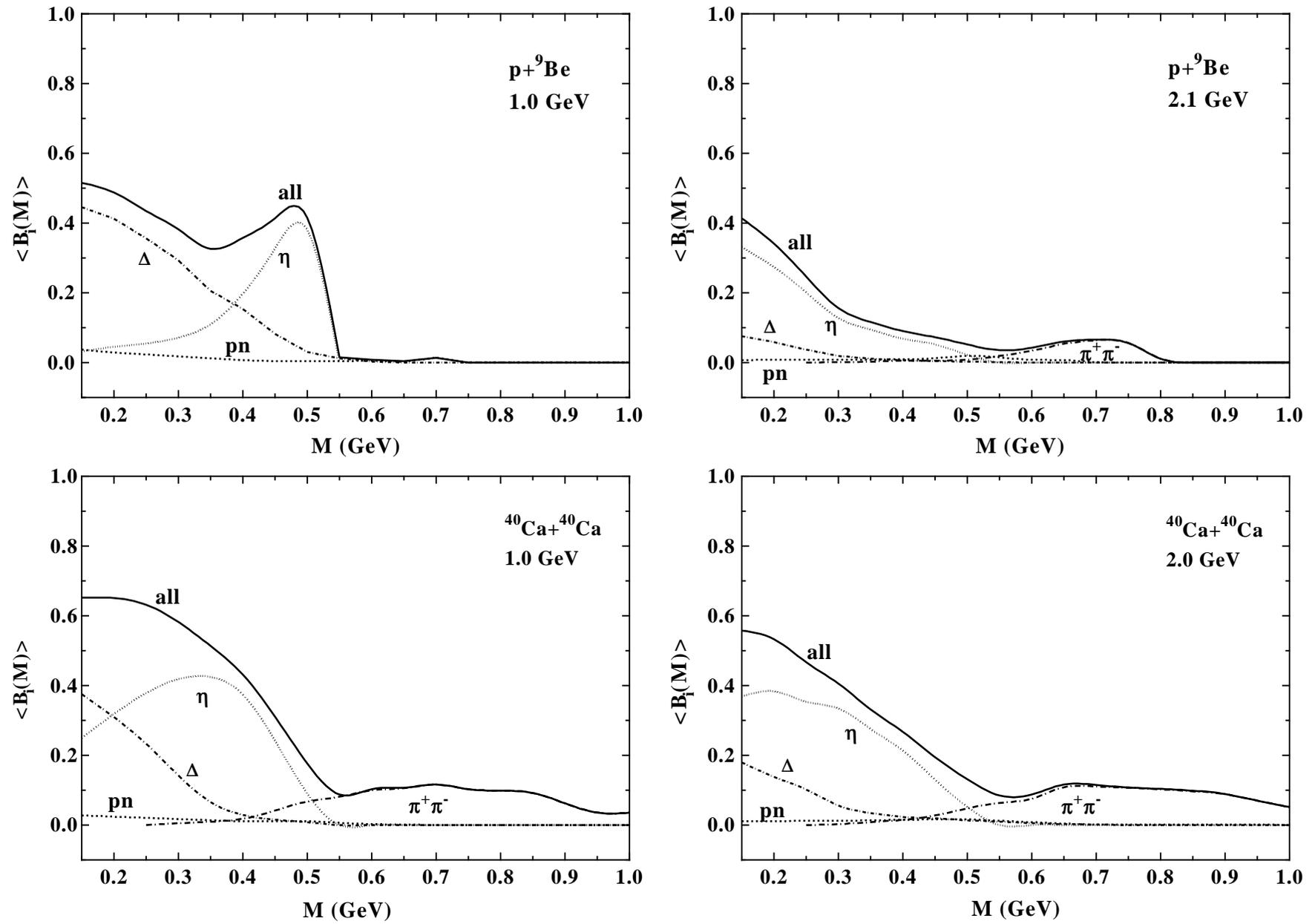
\end{document}